\documentstyle[prl,aps]{revtex}
\draft
\begin{document}
\input{epsf}
\twocolumn[\hsize\textwidth\columnwidth\hsize\csname@twocolumnfalse\endcsname
\title{Regularization of Synchronized Chaotic Bursts}
\author{Nikolai F. Rulkov}
\address{
Institute for Nonlinear Science, University of California, San
Diego, La Jolla, CA 92093-0402\\}

\date{\today}
\maketitle
\begin{abstract}
The onset of regular bursts in a group of irregularly bursting
neurons with different individual properties is one of the most
interesting dynamical properties found in neurobiological systems.
In this paper we show how synchronization among chaotically
bursting cells can lead to the onset of regular bursting. In order
to clearly present the mechanism behind such regularization we
model the individual dynamics of each cell with a simple
two-dimensional map that produces chaotic bursting behavior similar
to biological neurons.

\end{abstract}

\narrowtext
\vskip1pc]

Studies of cooperative behavior in coupled chaotic oscillators are
frequently based upon the analysis of the phenomenon of chaos
synchronization~\cite{synchr}. Different types of synchrony between
chaotic oscillators along with the various mechanisms responsible
for the onset of such synchronization have been
studied~\cite{sync}. These studies are mostly motivated by the
development of a theoretical framework that can explain
synchronization in neurobiological systems where complex, chaotic
behavior of neurons is quite typical~\cite{biosynch}.

It has been observed in neurobiological experiments and in
numerical simulations that individual neurons may show irregular
bursts, while ensembles of such irregularly bursting neurons can
synchronize and produce regular, rhythmical bursting. This
regularization occurs despite the significant differences in the
individual dynamics of coupled neurons~\cite{CPG}. How can a group
of different neurons, whose individual dynamics are very sensitive
to the intrinsic parameters of the neuron, synchronize and produce
a robust rhythm of the bursts? How can such rhythmical bursting
become insensitive to the intrinsic parameters that control the
individual dynamics of the neurons? Answers to these questions may
be found through the study of models, which, on one hand, are
capable of demonstrating similar effects and, on the other hand,
are simple enough to reveal the mechanisms that lead to these
effects.

\begin{figure}
\begin{center}
\leavevmode
\hbox{%
\epsfxsize=7.5cm
\epsffile{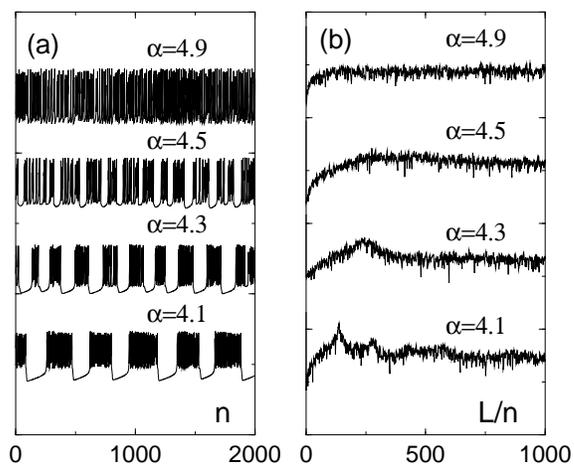}}
\end{center}
\caption{The waveforms (left) and power spectra (right) of the chaotic
behavior of uncoupled cells, $\epsilon=0$, computed for different
values of $\alpha$, with $\sigma=\beta=0.001$. The power spectra
are computed for waveforms of length $L$, ($L$=50000 iterations).}
\label{fig1}
\end{figure}

In this communication we present an example that demonstrates the
effects of mutual synchronization and {\em chaos regularization} of
bursts in a group of chaotically bursting cells. We study $N$
oscillators, which are modeled with two-dimensional maps and
coupled to each other through the mean field~\cite{coupling}
\begin{eqnarray}\label{maps}
x(i,n+1)&=&\frac{\alpha}{(1+x(i,n)^2)}+y(i,n)+\frac{\epsilon}{N}\sum_{j=1}^{N}x(j,n)~,
\nonumber \\
y(i,n+1)&=&y(i,n)-\sigma x(i,n)-\beta~, \label{nmaps}
\end{eqnarray}
where $x(i,n)$ and $y(i,n)$ are, respectively, the fast and slow
dynamical variables of the $i$-th oscillator, $\epsilon$ is the
strength of global coupling. The slow evolution of $y(i,n)$ is due
to the small values of the positive parameters $\beta$ and
$\sigma$, which are on the order of 0.001. The value of parameter
$\alpha$ is selected in the region $\alpha>4.0$, where the map
produces chaotic oscillations in $x(i,n)$. When $\epsilon=0$,
depending on the value of parameter $\alpha$, each cell
demonstrates two qualitatively different regimes of chaotic
behavior: continuous chaotic oscillations and chaotic bursts, see
Fig.~\ref{fig1}a. Such dynamics resemble the bursting dynamics
measured in the experiments with biological neurons, see for
example~\cite{elson98}.

\begin{figure}
\begin{center}
\leavevmode
\hbox{%
\epsfxsize=7.5cm
\epsffile{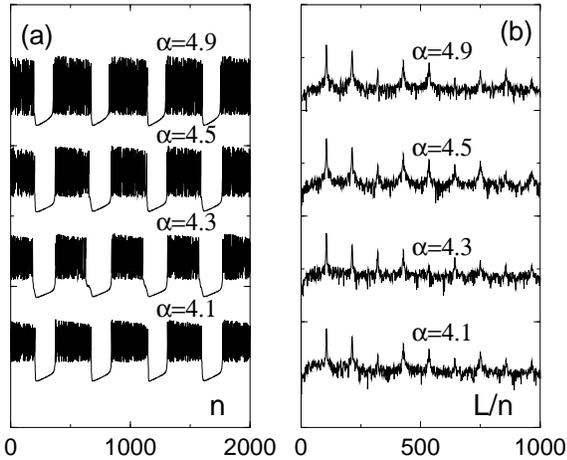}}
\end{center}
\caption{The waveforms (left) and power spectra (right) of the
synchronized chaotic bursts of the coupled cells (\ref{nmaps}) with
$\epsilon=0.2$ and $N=256$. The values of the individual parameters
of the cells are randomly selected from the intervals: $\alpha$
from 4.1 to 4.9; $\beta$ and $\sigma$ from 0.0009 to 0.0011. The
value of $L$ is the same as in Fig.~\ref{fig1}}
\label{fig2}
\end{figure}

As one can see from Fig.~\ref{fig1}a, the mean duration of the
bursts is very sensitive to the value of $\alpha$. The chaotic
nature of the duration of the bursts can be clearly seen from the
power spectra calculated from $x(i,n)$. The low-frequency part of
the spectrum that corresponds to the frequency range of the bursts
is presented in Fig.~\ref{fig1}b. The introduction of coupling
between the cells results in the synchronization of the bursts.
While the bursts do become synchronized, the fast chaotic
oscillations of the cells during the bursts remain asynchronous.
The most interesting effect observed in this case is that the
synchronized bursts measured both in each chaotic cell and in the
mean field ($\sum_{j=1}^{N}x(j,n)/N$) become almost periodic, see
Fig~\ref{fig2}a. We define the formation of such a periodic
temporal pattern in the chaotic behavior of systems as {\em chaos
regularization}. The effect of chaos regularization can be clearly
seen in the power spectrum of $x(i,n)$ as the appearance of
periodic spectral components (compare Fig.~\ref{fig1}b and
Fig.~\ref{fig2}b). These periodic components correspond to the
period and the harmonics of the periodic temporal pattern formed in
the synchronized chaotic bursts. In the rest of the paper we will
study the mechanisms behind the chaos regularization of the bursts
in the model cells (\ref{nmaps}).

{\em Chaotic Bursting}. The individual behavior of the cells is
described by the map of the form
\begin{eqnarray}
x_{n+1}&=&\frac{\alpha}{(1+x_n^2)}+y_n~,
\label{umapx} \\
y_{n+1}&=&y_n-\sigma x_n-\beta~. \label{umapy}
\end{eqnarray}
The slow evolution of $y_n$ for the next $m$ steps is given by
\begin{eqnarray}
y_{n+m}&=&y_n-m(\beta + \sigma \overline{x_{n,m}}), \label{mapy}
\end{eqnarray}
where $\overline{x_{n,m}}= (\sum_{j=n+1}^{n+m} x_j)/m$ is the mean
value of $x_n$ computed for $m$ consecutive iterations. It is clear
from (\ref{mapy}) that the value of $y_n$ slowly increases during
the next $m$ steps if $\sigma \overline{x_{n,m}} < -\beta$, and
decreases if $\sigma \overline{x_{n,m}} > -\beta$.

\begin{figure}
\begin{center}
\leavevmode
\hbox{%
\epsfxsize=4.2cm
\epsffile{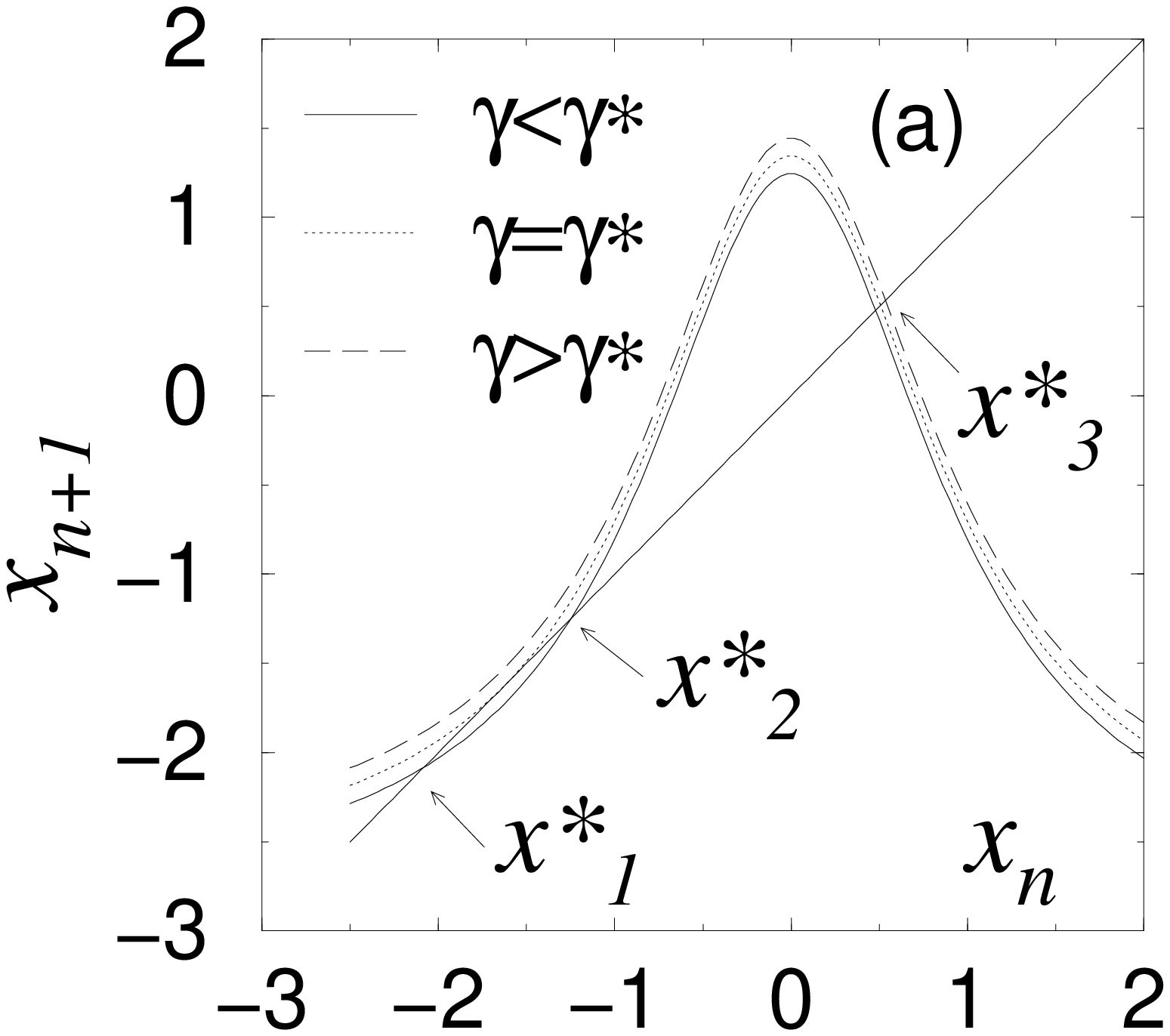}
\epsfxsize=4.2cm
\epsffile{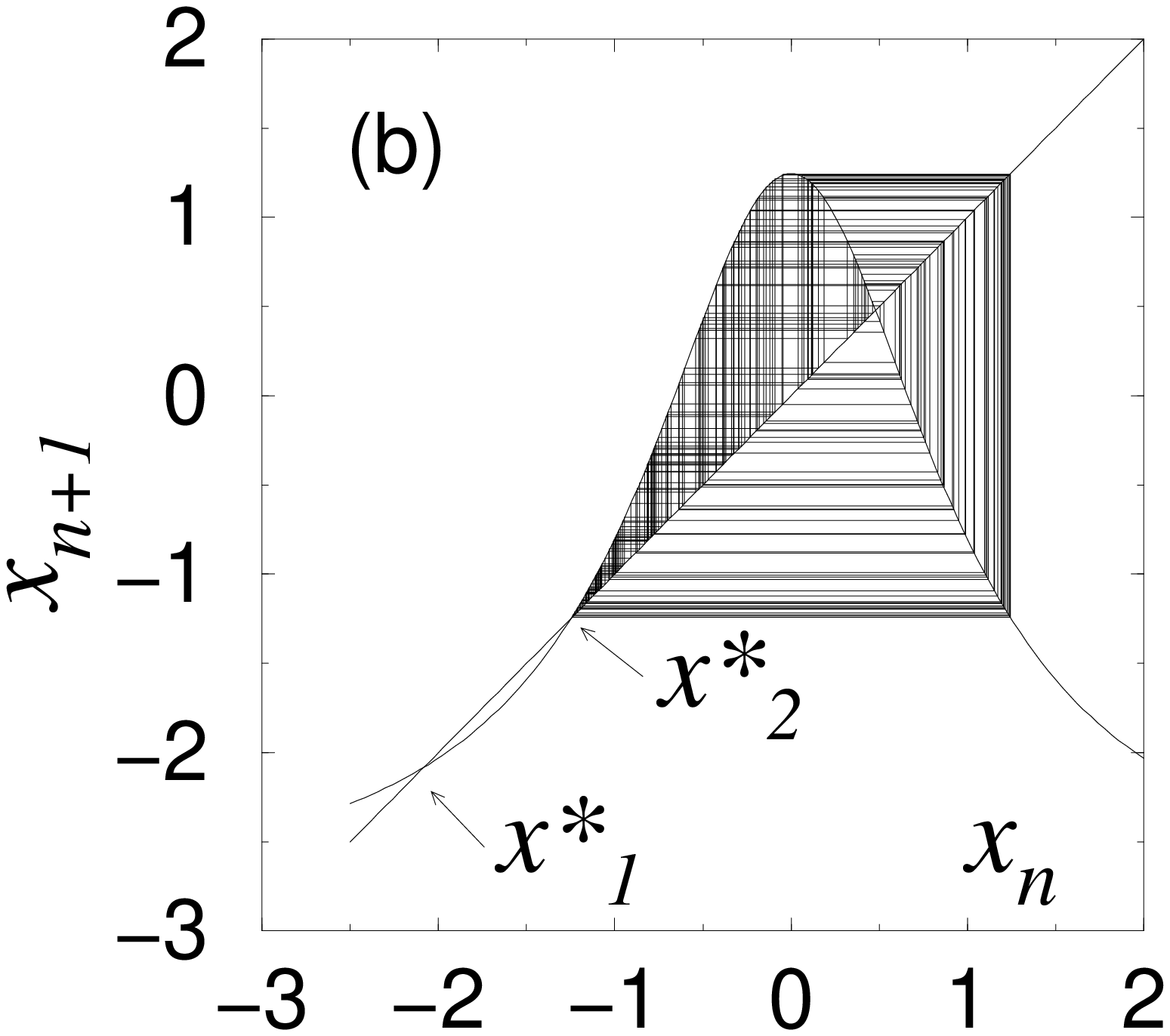}}
\end{center}
\caption{The shape of the function $F(x_n,\alpha,\gamma)$ plotted for
$\alpha=4.1$ and three different values of $\gamma$ (-2.65, -2.75
and -2.85) (a) and the chaotic trajectory of the map (b). $x^*_1$,
$x^*_2$ and $x^*_3$ are the fixed points of the map. $\gamma^*$
corresponds to the bifurcation value where $x^*_1$ and $x^*_2$
merge together.}
\label{mplot}
\end{figure}

\begin{figure}
\begin{center}
\leavevmode
\hbox{%
\epsfxsize=3.9cm
\epsffile{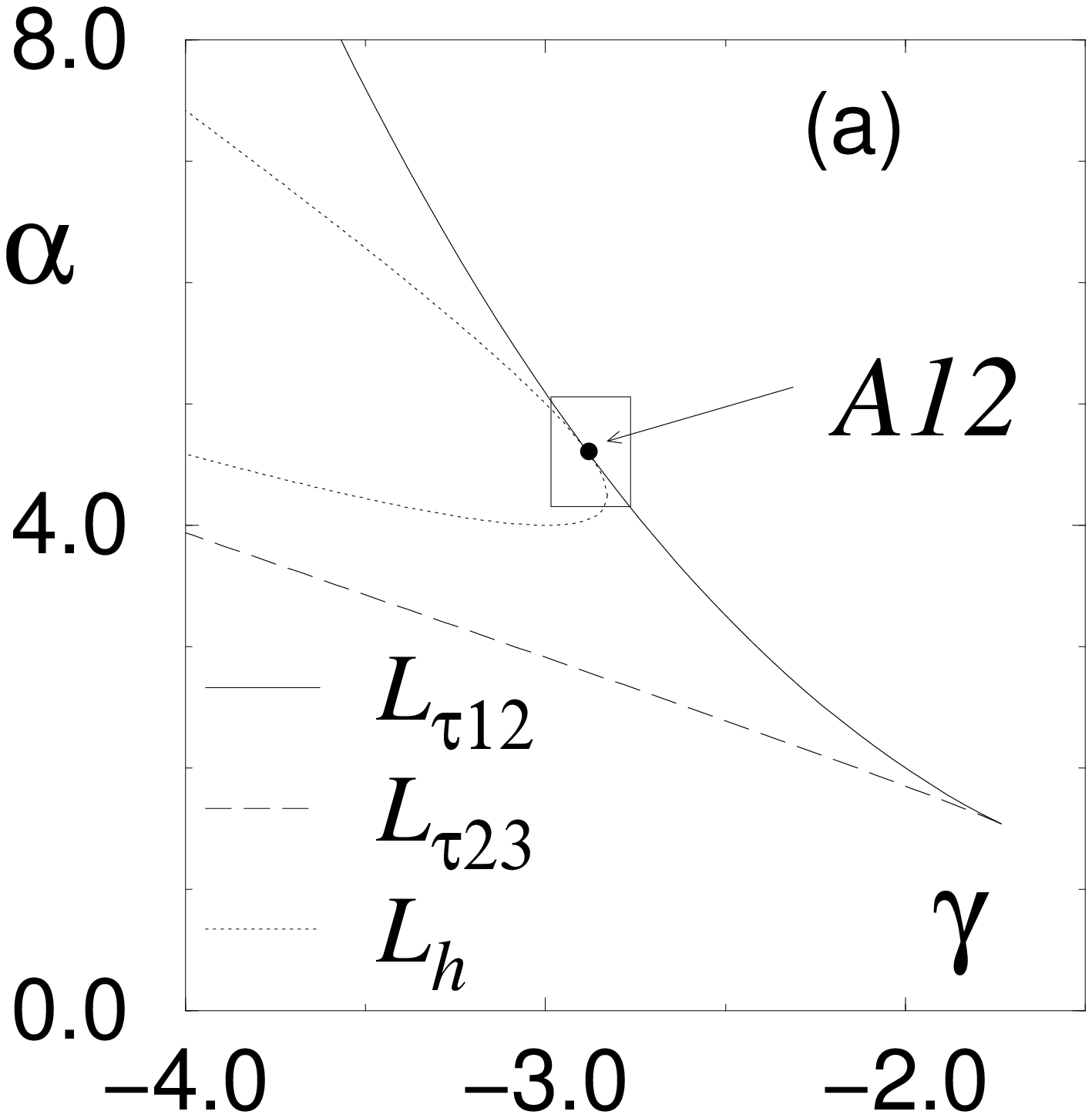}
\epsfxsize=3.9cm
\epsffile{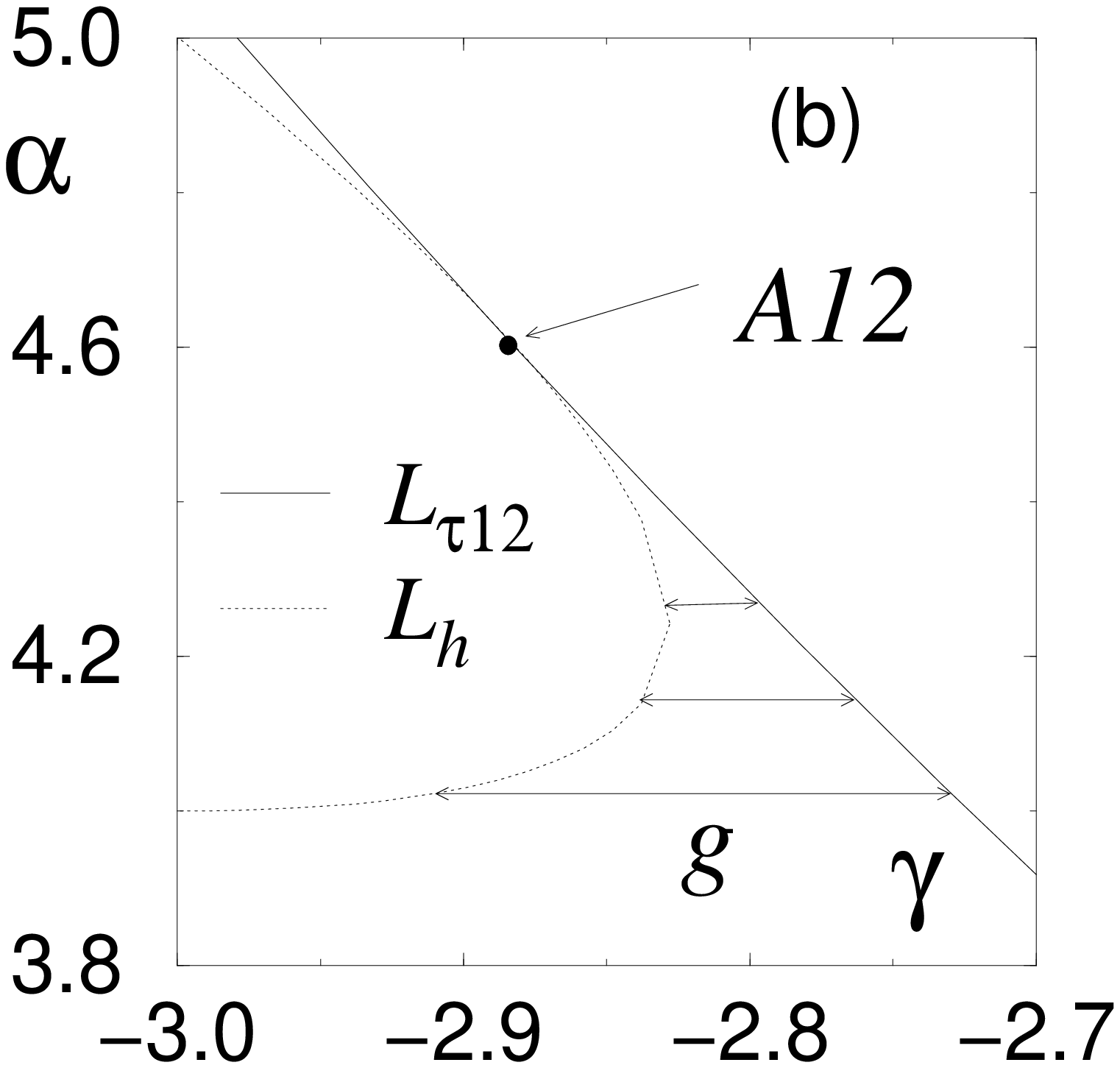}}
\end{center}
\caption{The bifurcation diagram of the uncoupled fast map (a), and
the enlarge portion of the diagram around the point $A12$ (b). The
bifurcation curves corresponding to the merging of the fixed points
$x^*_j$ and $x^*_m$ (curves $L_{\tau jm})$ and to the crisis of the
chaotic attractor (curve $L_h$). $g$ shows the gap between the
bifurcation curves.}
\label{mbif}
\end{figure}

Since $y_n$ changes slowly, the dynamics of $x_n$ can be considered
independently of the map (\ref{umapy}) assuming that $y_n$ is a
parameter $\gamma = y_n$. Therefore, fast dynamics of the cell can
be understood from the analysis of the 1-D map
\begin{eqnarray}
x_{n+1}&=&F(x_n,\alpha,\gamma)\equiv\frac{\alpha}{(1+x_n^2)}+\gamma~.
\label{mapx}
\end{eqnarray}
The main dynamical properties of this map are clear
from~Fig~\ref{mplot}. The results of the bifurcation analysis of
the map are presented in Fig.~\ref{mbif}. If the parameter values
of the map are within the horn formed by the bifurcation curves
$L_{\tau12}$ and $L_{\tau23}$, given by the equations
$\alpha=-2((\gamma^2+9)\gamma-(\gamma^2-3)^{3/2})/27$ and
$\alpha=-2((\gamma^2+9)\gamma+(\gamma^2-3)^{3/2})/27$ respectively,
then the map~(\ref{mapx}) has three fixed points $x^*_1$, $x^*_2$
and $x^*_3$. The fixed point $x^*_1$ is stable, $x^*_2$ is unstable
and $x^*_3$ may change stability, see Fig.~\ref{mplot}.

When the parameter values of the map cross the curve $L_{\tau12}$
from left to right then the fixed points $x^*_1$ and $x^*_2$ merge
together and disappear. This bifurcation in (\ref{mapx})
corresponds to the beginning of the burst in system (\ref{umapx})
and (\ref{umapy}). Indeed, when $x_n$ is in the stable fixed point
$x^*_1$, then $\gamma
\equiv y_n$ slowly grows because $\overline{x_{n,1}}\approx x^*_1 <
-\beta/\sigma$, see~(\ref{mapy}). When the stable fixed point $x^*_1$
disappears, the trajectory of the fast map~(\ref{mapx}) goes to the
chaotic attractor corresponding to the chaotic pulsations during
the burst, see Fig.~\ref{mplot}.

The end of the chaotic burst is due to the external crises of the
chaotic attractor in system (\ref{mapx}). When the trajectory of
this map is on the chaotic attractor, then
$\overline{x_{n,1}}>-\beta/\sigma$ (see Fig.~\ref{mplot}), and the
value of $\gamma \equiv y_n$ decreases until the chaotic trajectory
of the fast map arrives back to the stable fixed point $x^*_1$.
This becomes possible only after the parameter of the fast map goes
through the bifurcation value, which corresponds to the case when
the trajectory from the maximum of the map function
$x_{max}=F(0,\alpha,\gamma)$ maps into the unstable fixed point
$x^*_2$. Curve $L_h$, that corresponds to the bifurcation values
where $x_{max}$ iterates into a fixed point $x^*_2$ or $x^*_1$, is
given by the equation $\alpha=-(3\gamma
\pm\sqrt{\gamma^2-8})/2 $ and is shown in Fig.~\ref{mbif} . Note that
the external crises of the chaotic attractor is associated only
with the fixed point $x^*_2$. This bifurcation corresponds to the
low branch of the curve $L_h$ that starts from point $A12$.

Therefore, the duration of the chaotic burst is determined by the
time interval that is required for the slow variable $y_n$ to move
from the bifurcation curve $L_{\tau12}$ to the curve $L_h$ plus the
additional time interval that is needed for the trajectory to map
below the unstable fixed point $x^*_2$. Both of these intervals
fluctuate because of the chaotic nature of the trajectory $x_n$,
and this is the reason for the chaotic fluctuations of the duration
of the bursts. After the chaotic firing terminates the trajectory
arrives at the stable fixed point $x^*_1$, and $\gamma$ slowly
moves towards $L_{\tau12}$, where the next burst starts. One can
see from Fig~\ref{mbif}b that the size of gap $g$ between the
bifurcation curves $L_{\tau12}$ and $L_h$ is very sensitive to the
value of parameter $\alpha$. That explains the high sensitivity of
the mean duration of the bursts to the internal parameter of the
cell $\alpha$.

If the value of parameter $\alpha$ is selected above the point
$A12$, then the fixed points $x^*_1$ and $x^*_2$ appear within the
chaotic attractor (internal crisis). In this case $y_n$ fluctuates
in the vicinity of the bifurcation curve $L_{\tau 12}$ keeping the
cell in the regime of continuous chaotic oscillations, see
Fig.~\ref{fig1}a. We would like to note that the considered
mechanism of chaotic bursting is essentially the same as in the
Hindmarsh-Rose model of biological neuron, where the role of
parameter $\alpha$ is played by a hyperpolarization current, see
Ref.~\cite{HRModel}.

{\em Synchronization}. Coupling between the cells (\ref{nmaps})
influences the fast dynamics of each cell (\ref{mapx}) by adding
the value $\epsilon \sum_{j=1}^{N}x(j,n)/N$ to the parameter
$\gamma$. When the $i$-th cell approaches the bifurcation values at
$L_{\tau12}$, being in the stable fixed point $x^*_1$, its behavior
becomes very sensitive to the influence of the other cells. The
cells, which are crossing this curve and starting to fire, sharply
increase their values of $x(j,n)$. As a result, the increased value
of $\gamma$ in the $i$-th cell pushes the cell over the bifurcation
value and triggers the burst. The greater the number of cells which
are involved in the simultaneous transition to the firing phase,
the larger the triggering impact will be that is experienced by the
remaining cell. Such an avalanche type of synchronous switching
takes place both at the beginning and at the end of each burst.
This threshold mechanism of synchronization is similar to the one
studied in ensembles of integrate-and-fire
oscillators~\cite{strogatz}.

\begin{figure}
\begin{center}
\leavevmode
\hbox{%
\epsfxsize=4.65cm
\epsffile{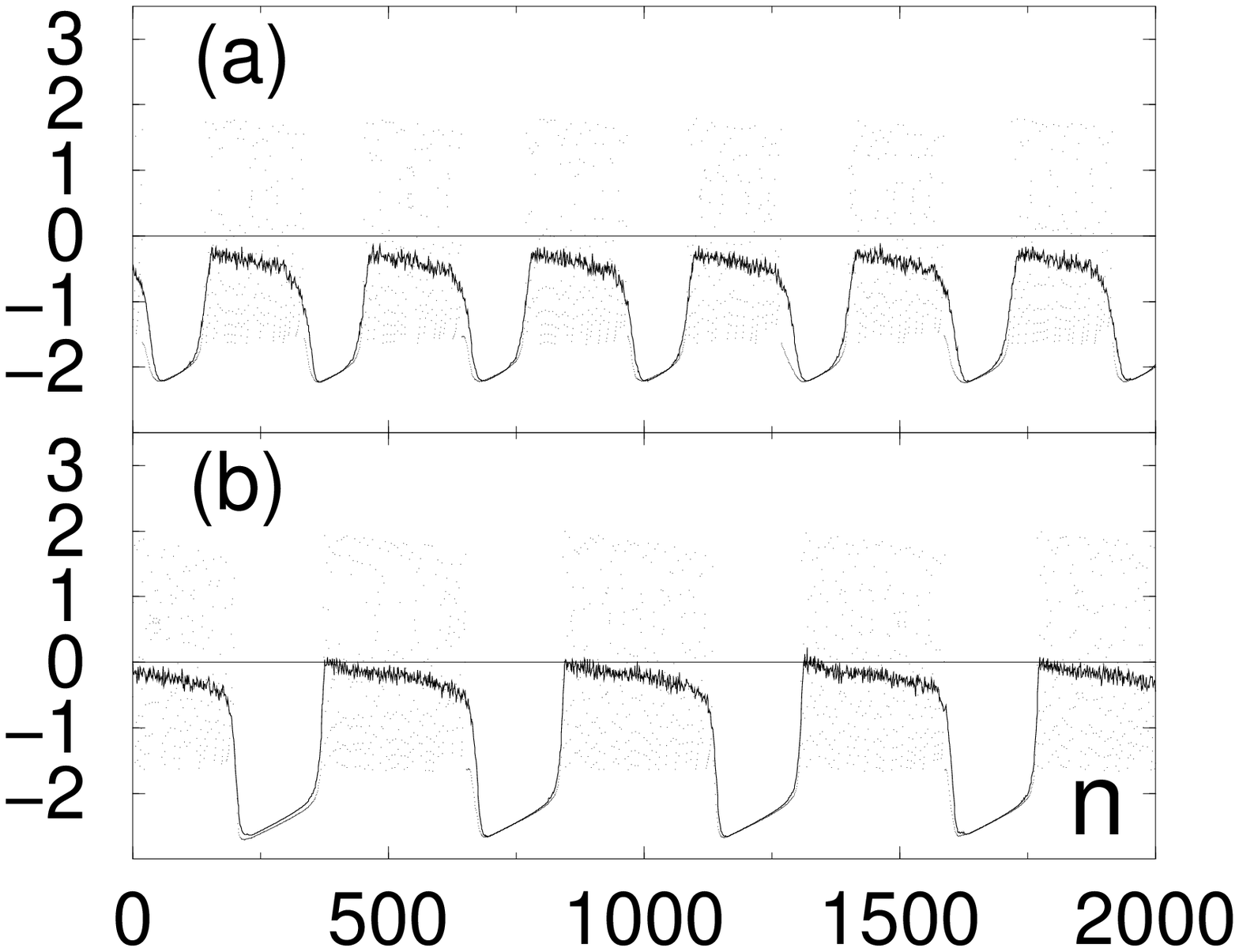}
\epsfxsize=3.5cm
\epsffile{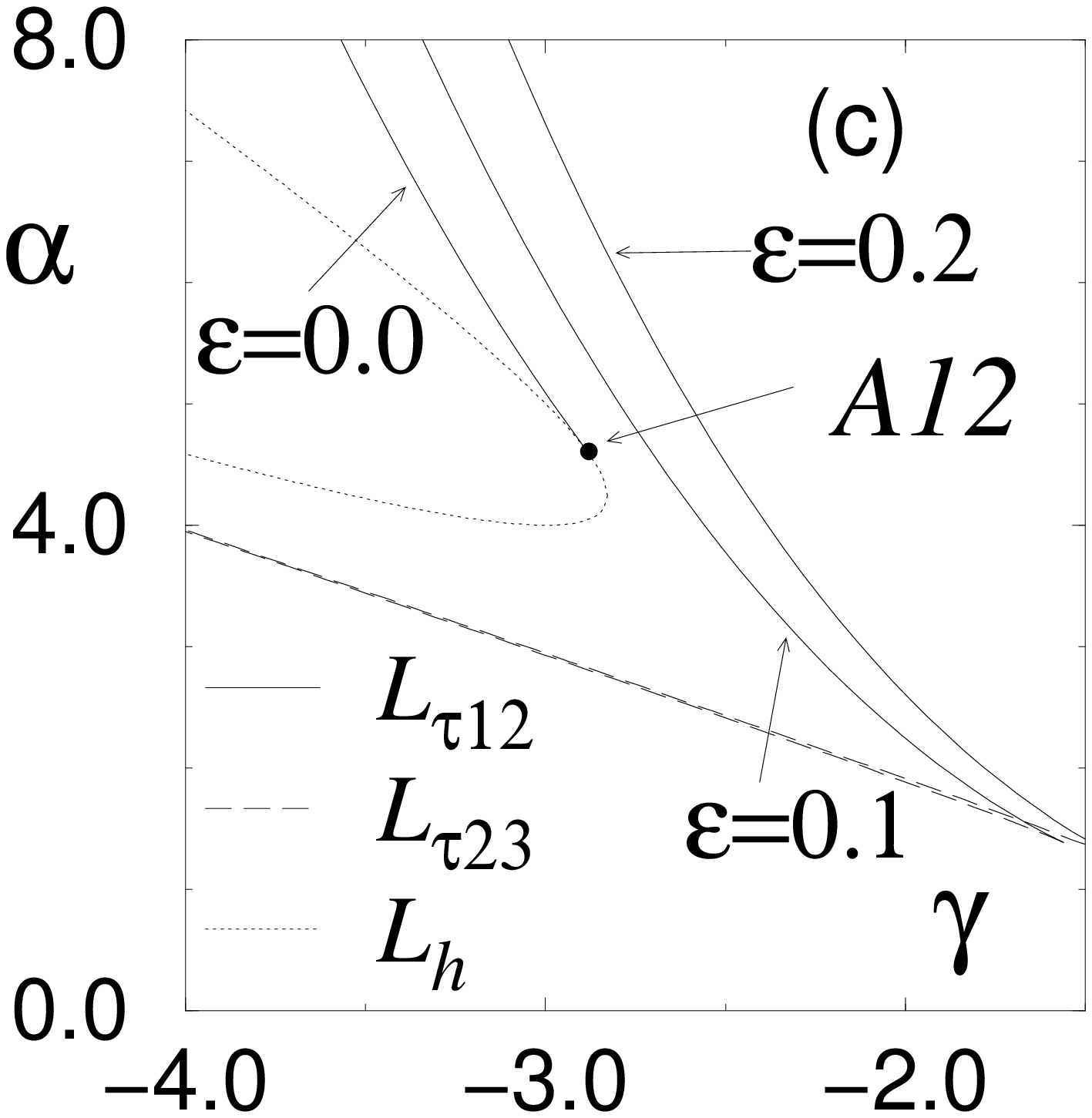}}
\end{center}
\caption{The waveforms of the mean field (solid line) and of $x(i,n)$
for the cell with $\alpha=4.5$ (dots) computed for the group of
synchronously bursting cells with $\epsilon=0.1$ (a) and
$\epsilon=0.2$ (b). The bifurcation curves of the fast maps in the
case of synchronously bursting cells (c).}
\label{mbifc}
\end{figure}

{\em Regularization} The dynamical mechanism of regularization can
be also understood from the bifurcations of the fast map
(\ref{mapx}) taking into account the alternation of the two phases
in the waveform of the mean field $\epsilon \sum_{j=1}^{N}x(j,n)/N$
for synchronized bursts, see Fig.~\ref{mbifc}a,b. When the
synchronously bursting cells are in the non-firing phase, the fast
variable of each cell is located in the stable fixed point $x^*_1$
and the values of $y(i,n)$ and the $x$-coordinates of the fixed
point $x^*_1(i,n)$ slowly increase. In this case the evolution of
$x$ in each cell can be described by the map (\ref{nmaps}), where
the coupling term can be approximated by the value $\epsilon
x(i,n)$.
\begin{eqnarray}
x(i,n+1)&=&\frac{\alpha}{(1+x(i,n)^2)}+
\gamma+\epsilon x(i,n).
\label{mapix}
\end{eqnarray}
The bifurcation curve $L_{\tau12}$, which defines the threshold
value of $y(i,n)$ corresponding to the beginning of the burst, is
now given by the equation
$27\alpha(1-\epsilon)^2/2=-(\gamma^2+9(1-\epsilon)^2)
\gamma-(\gamma^2- 3(1-\epsilon)^2)^{3/2}$.
The location of this curve on the parameter plane $(\gamma,\alpha)$
depends upon the value of the coupling parameter $\epsilon$ and
shifts to the right from the original position of $L_{\tau12}$, see
Fig.~\ref{mbifc}c.

When all synchronously bursting cells are in the phase of chaotic
firing their fast chaotic pulsations are uncorrelated. Therefore,
the value of the coupling term ($\epsilon \sum_{j=1}^{N}x(j,n)/N$)
during the burst can be approximated with the mean value of
$<x(j,n)>$. Due to the averaging, the fast chaotic fluctuation in
the coupling term disappears and the value $|\epsilon <x(j,n)>|$
slowly drifts in the interval between 0 and 0.08, see
Figs~\ref{mbifc} a,b. These values are significantly less than the
amplitude of the slow variable $|y_n|$, which is of the order of
one. Therefore, for simplicity, the contribution from the mean
field to the parameter $\gamma$ can be neglected during the firing
phase. As a result, the approximate position of the bifurcation
curves for the fast chaotic oscillations of synchronously firing
cells (\ref{nmaps}) can considered to be the same as for the
uncoupled map (\ref{mapx}).

Fig.~\ref{mbifc}c presents the bifurcation curve $L_{\tau12}$
($\epsilon>0$) that defines the beginning of a burst, and the
bifurcation curves $L_h$ (the part below $A12$) and $L_{\tau12}$
($\epsilon=0$, the part above $A12$), which define the end of the
burst in each synchronously bursting cell. It follows from this
figure that the states of slow variable $y(i,n)$ corresponding to
the beginning and the end of the bursts become separated for all
values $\alpha$. This separation defines the period of the periodic
pattern formed in the chaotically bursting cells and helps to
synchronize cells with qualitatively different individual dynamics
whose bursting rates become insensitive to the parameter $\alpha$.

Assuming that variables $y(i,n)$ drift with approximately the same
rate, the mean period of bursts, $T_B$, should be proportional to
the length of the most narrow gap $g$ between these two curves. One
can see from Fig.\ref{mbifc}c that the size of such a minimal gap
$g_{min}$ increases with the strength of the coupling. This
explains the change of bursts frequency in the waveforms shown in
Fig.\ref{mbifc}a,b.

The presented analysis shows that chaos regularization in a group
of chaotically bursting cells has the following mechanism. When the
large enough fraction of the group of cells starts to burst
simultaneously due to the threshold synchronization, it leads to
the formation of relatively sharp transitions (ramps) in the mean
field (coupling term). These ramps create a gap between the states
of $y(i,n)$ corresponding to the beginning and the end of the
bursts for the rest of the cells, including the continuously firing
ones. All cells become bursting cells and it improves the
synchronization among them. Due to the formation of the gap, a
large portion of the group have about the same mean period which is
short enough to lead the bursting of the whole ensemble, see
Fig.\ref{mbifc}c. During the chaotic burst the chaotic components
of mean field are small because of averaging over the whole group
of the cell and, as the result, these components do not influence
the individual oscillations of the cell. When a large number of
firing cells get close to the threshold, the leading cells trigger
an avalanche of transitions. This avalanche sharply switches the
mean field to the non-firing phase and the rest of the cells are
forced to change their state independently of their chaotic
trajectory. This is how the synchronization makes the duration of
the bursts insensitive to the chaotic trajectory. It is clear that
the regularity of the bursts improves when the number of cells
increases. However, due to the formation of the gap, this type of
chaos regularization becomes noticeable even for small groups of
cells, including the case $N=2$.

Finally, we would like to emphasize that this mechanism of chaos
regularization is due to the dynamical features of each cell at the
beginning and at the end of the chaotic burst. In the considered
map these features are characterized by bifurcations at
$L_{\tau12}$ and $L_h$, which are similar to a saddle-node
bifurcation and the appearance of a homoclinic orbit in more
realistic models of biological neurons, see for
example~\cite{HRModel}. Since the mechanism of chaos regularization
relies only upon the formation of sharp ramps in the coupling
forces, and the averaging down of the chaotic fluctuation during
the firing phase, this mechanism can occur for a number of
different types of coupling among the cells.

The author is grateful to R. Elson, M.I. Rabinovich and H.D.I.
Abarbanel for helpful discussions. This work was supported in part
by U.S. Department of Energy (grant DE-FG03-95ER14516) and the U.S.
Army Research Office (MURI grant DAAG55-98-1-0269).

\end{document}